\documentclass[aps,prd,twocolumn,preprintnumbers,groupedaddress,showpacs,nofootinbib,amssymb]{revtex4}
\usepackage[T1]{fontenc}
\usepackage[latin1]{inputenc}
\usepackage{graphicx}
\usepackage[english]{babel}
\usepackage{amsmath}
\usepackage{amssymb}
\usepackage{amsfonts}

\begin{document}

\def\pp{{\, \mid \hskip -1.5mm =}}
\def\cL{{\cal L}}
\def\be{\begin{equation}}
\def\ee{\end{equation}}
\def\bea{\begin{eqnarray}}
\def\eea{\end{eqnarray}}
\def\tr{{\rm tr}\, }
\def\nn{\nonumber \\}
\def\e{\mathrm{e}}
\def\ki{{\cal X}} 
\def\L{{\cal L}} 
\def\F{{\cal F}} 
\def\1{{\bold 1}}
\def\LG{L_{\Gamma}}  
\def\G{\Gamma}  
\def\d{\partial}    
\def\alfa{\alpha}    
\def\R{{\bf R}}  
\def\C{\bf C}
\def\eqdef{\buildrel {\rm def} \over =}     
\def\({\left(}    \def\){\right)}
\def\est{\wedge}      
\def\diff{diffeomorphism}
\def\colon{\quad;\quad}    
\def\Colon{\qquad;\qquad}
\def\teta{\theta}   
\def\tL{\theta_\L}
\def\data{\the\day-\the\month-\the\year}
\def\toptitle#1{\voffset=2\baselineskip 
                \headline={\it #1 \hrulefill \ \ \data} 
                }
\def\bottomtitle#1{
\footline={#1 \hskip5pt \data \hfill \folio}
                  }
\def\hdot{{\dot H}}
\def\adot{{\dot a}}  
\def\phidot{{\dot \phi}} 
\def\addot{{\ddot a}}
\def\phiddot{{\ddot \phi}}   
\def\LL{\Lambda}  
\def\D{\Delta}
\def\frac#1#2{{#1 \over #2}}
\def\newsec#1{\vskip20pt {\bf \noindent #1} \vskip10pt}
\def\comment#1{{\bf #1}}
\def\phi{\varphi} 
\def\M{{\scriptscriptstyle M}} 
\def\T{{\scriptscriptstyle T}}
\def\ro{\rho}
\def\rom{\rho_\M} 
\def\PM{P_\M}
\def\rot{\rho_\T} 
\def\rof{\rho_\phi} 
\def\PT{P_\T} 
\def\PF{P_\phi}
\def\gam{\gamma_\M} 
\def\gaf{\gamma_\phi} 
\def\gat{\gamma_\T}
\def\half{\frac{1}{2}} 
\def\~{\approx}
\def\beqa{\begin{eqnarray}}
\def\eeqa{\end{eqnarray}}
\def\beq{\begin{equation}}
\def\eeq{\end{equation}}

\def\gb{\bar{g}}
\def\ad{\dot{a}}
\def\vol{\int d^4x\,\sqrt{-g}}
\def\grav{\frac{1}{16 \pi G}}
\def\half{\frac{1}{2}}
\def\gu{g^{\mu\nu}}
\def\gd{g_{\mu\nu}}
\def\p{\phi}
\def\v{V(\phi)}
\def\vp{V'(\phi)}
\def\l{\cal L}
\def\bib#1{[{\ref{#1}}]}
\def\mps{{\bf \Psi}}
\def\bp{{\bf \psi}}
\def\t{\cal T}
\def\ad{\dot{a}}
\def\vol{\int d^4x\,\sqrt{-g}}
\def\grav{\frac{1}{16 \pi G}}
\def\half{\frac{1}{2}}
\def\gu{g^{\mu\nu}}
\def\gd{g_{\mu\nu}}
\def\umu{^{\mu}}
\def\unu{^{\nu}}
\def\dmu{_{\mu}}
\def\dnu{_{\nu}}
\def\umunu{^{\mu\nu}}
\def\dmunu{_{\mu\nu}}
\def\ua{^{\alpha}}
\def\ub{^{\beta}}
\def\da{_{\alpha}}
\def\db{_{\beta}}
\def\ug{^{\gamma}}
\def\dg{_{\gamma}}
\def\uamu{^{\alpha\mu}}
\def\uanu{^{\alpha\nu}}
\def\uab{^{\alpha\beta}}
\def\dab{_{\alpha\beta}}
\def\dabgd{_{\alpha\beta\gamma\delta}}
\def\uabgd{^{\alpha\beta\gamma\delta}}
\def\udeab{^{;\alpha\beta}}
\def\ddeab{_{;\alpha\beta}}
\def\ddemunu{_{;\mu\nu}}
\def\udemunu{^{;\mu\nu}}
\def\ddemu{_{;\mu}}  \def\udemu{^{;\mu}}
\def\ddenu{_{;\nu}}  \def\udenu{^{;\nu}}
\def\ddea{_{;\alpha}}  \def\udea{^{;\alpha}}
\def\ddeb{_{;\beta}}  \def\udeb{^{;\beta}}
\def\naba{\nabla_{\alpha}}
\def\nabb{\nabla_{\beta}}
\def\pnu{\partial_{\nu}}
\def\pa{\partial}
\def\bib#1{$^{\ref{#1}}$}
\def\lam{\lambda}
\def\Lam{\Lambda}
\def\eps{\varepsilon}
\def\gam{\gamma}
\def\alp{\alpha}
\def\sig{\sigma}
\def\bq{{\bf q}}
\def\bqd{{\bf \dot{q}}}
\def\qd{\dot{q}}

\def\ki{{\cal X}}
\def\L{{\cal L}}
\def\F{{\cal F}}
\def\LG{L_{\Gamma}}
\def\G{\Gamma}
\def\d{\partial}
\def\alfa{\alpha}
\def\R{{\bf R}}
\def\C{\bf C}
\def\eqdef{\buildrel {\rm def} \over =}
\def\({\left(}    \def\){\right)}
\def\est{\wedge}
\def\diff{diffeomorphism}
\def\colon{\quad;\quad}
\def\Colon{\qquad;\qquad}
\def\teta{\theta}
\def\tL{\theta_\L}
\def\data{\the\day-\the\month-\the\year}
\def\toptitle#1{\voffset=2\baselineskip
                \headline={\it #1 \hrulefill \ \ \data}
                }
\def\bottomtitle#1{
\footline={#1 \hskip5pt \data \hfill \folio}
                  }
\def\hdot{{\dot H}}
\def\adot{{\dot a}}
\def\phidot{{\dot \phi}}
\def\addot{{\ddot a}}
\def\phiddot{{\ddot \phi}}
\def\LL{\Lambda}
\def\D{\Delta}
\def\frac#1#2{{#1 \over #2}}
\def\newsec#1{\vskip20pt {\bf \noindent #1} \vskip10pt}
\def\comment#1{{\bf #1}}
\def\phi{\varphi}
\def\M{{\scriptscriptstyle M}}
\def\T{{\scriptscriptstyle T}}
\def\ro{\rho}
\def\rom{\rho_\M}
\def\PM{P_\M}
\def\rot{\rho_\T}
\def\rof{\rho_\phi}
\def\PT{P_\T}
\def\PF{P_\phi}
\def\gam{\gamma_\M}
\def\gaf{\gamma_\phi}
\def\gat{\gamma_\T}
\def\half{\frac{1}{2}}
\def\~{\approx}
\def\psiul{\overline\psi}
\def\pb{\not\!\partial}
\def\f{F(\phi)}
\def\fp{F'(\phi)}
\def\fpp{F''(\phi)}
\def\p{\phi}
\def\v{V(\phi)}
\def\vp{V'(\phi)}
\def\l{\cal L}
\def\bib#1{[{\ref{#1}}]}


\title{Gauss-Bonnet dark energy by Lagrange multipliers}

\author{Salvatore Capozziello$^{1,2}$, Andrey N. Makarenko $^4$  and Sergei D. Odintsov$^{3,4,5}$}

\affiliation{$^1$Dipartimento di Scienze Fisiche, Universit\`{a} di Napoli "Federico II", Napoli, Italy\\
$^2$INFN Sez. di Napoli, Compl. Univ. di Monte S. Angelo, Edificio G, Via Cinthia, I-80126, Napoli, Italy,\\
$^3$Instituci\`{o} Catalana de Recerca i Estudis Avan\c{c}ats
(ICREA) and Institut de Ciencies de l'Espai (IEEC-CSIC),Facultad de
Ciencies,C5, Campus UAB, 08193 Barcelona, Spain \\
$^4$ Tomsk State Pedagogical University, Tomsk, Russia \\
$^5$ Eurasian National University,
Astana, Kazakhstan}

\date{\today}

\begin{abstract}
A string-inspired effective theory of gravity, containing Gauss-Bonnet
invariant interacting with a scalar field, is considered in view of obtaining cosmological dark energy solutions. A  Lagrange multiplier is inserted into the action in order to achieve the cosmological reconstruction by selecting suitable forms of couplings and potentials. 
Several cosmological exact solutions (including dark energy of quintessence, 
phantom or Little Rip type) are derived in presence and in absence of the Lagrange multiplier showing the difference in the two dynamical approaches. 
In the models that we  consider, the Lagrange multiplier behaves as a sort of dust fluid  that  realizes  the  transitions between  matter dominated  and
dark energy epochs.  The relation between Lagrange multipliers and Noether symmetries is discussed.

\end{abstract}


\maketitle

\section{Introduction}

Astrophysical data indicate that the observed universe is  in an
accelerated phase \cite{Dat}. This acceleration is induced
by  the so-called dark energy
(see Ref. \cite{review} for a recent review and references therein) which nature and properties are
not yet understood at fundamental level.
In the most theoretical models considered in the literature, the dark energy is 
constituted  by
some ideal fluid with a specific equation of state (EoS) sometimes
exhibiting non-standard  properties  like  negative pressure and/or a 
negative entropy.
On the other hand, dark energy can  be considered as a global phenomenon associated with modifications of  
gravity   \cite{review1}. In fact,  its presence  could point out  
that Einstein's  General Relativity cannot be retained as the final theory of gravitational interaction at cosmological scales.
In this sense, General Relativity presents problems at UV (Quantum Gravity) and IR (cosmology) scales \cite{faraoni}. 
A similar situation appear for dark matter phenomena. No fundamental candidate has been revealed up to now and dynamics of self-gravitating structures could be addressed by modifications of gravity \cite{annalen}.

A further problem is  that  it is not clear why  dark energy had no effect at  early 
epochs
while it gives dominant contribution in  today observed  universe. According to the 
latest observational data, dark energy currently accounts for about 73\% of the total mass-energy amount of
the universe (see, for example, Ref.~\cite{Kowalski}).

The main feature of dark energy is that its EoS parameter  $w_\mathrm{D}$ is
negative:
\begin{equation}
w_\mathrm{D}=p_\mathrm{D}/\rho_\mathrm{D}<0\, , \end{equation} where
$\rho_\mathrm{D}$ is the dark energy density and $p_\mathrm{D}$ the pressure. According to the standard cosmological model, this property gives rise to the reported apparent acceleration of the Hubble fluid.
Although current data favor the standard $\Lambda$CDM cosmology,
the uncertainties in the determination of the EoS dark energy parameter $w$ are
still too large, namely $w=-1.04^{+0.09}_{-0.10}$. Hence, one is not able to 
determine,
without doubt, which of the three cases:
$w < -1$, $w = -1$, or $w >-1$ is the one actually realized in our universe
\cite{PDP,Amman}. Future observations should better constrain this range of values giving also indications on the   nature  of dark energy \cite{euclid}. 

In order to explain  dark energy at very fundamental level,   string/M-theory could  suggest reliable  
effective models to be compared with observations.
In particular,  further gravitational terms, emerging from this theory,  could become important at
current, low-curvature universe (being not essential at intermediate epochs
from strong to low curvature). For instance, in the study of
string-inspired gravity near the initial singularity, the role of the Gauss-Bonnet 
(GB) topological term, coupled
  with scalar potentials,   is important for
the occurrence of non-singular cosmology \cite{ART,nick}. The
 dilaton coupled to higher-order curvature  corrections, for example, assumes an important role  near the initial singularity, as discussed 
in  \cite{modulus}.
Furthermore, string-inspired gravity with
   Gauss-Bonnet term interacting with a scalar field 
has been proposed as a realistic candidate to address the  dark energy issue \cite{Odin1}.
Specifically, as it was shown in Ref.\cite{Odin1},   Gauss-Bonnet dark energy may lead to the occurrence of 
phantom cosmology  without ghosts: in this case, the  dilaton is a  canonical 
scalar. Further aspects of Gauss-Bonnet accelerating cosmology
have been 
discussed in detail in \cite{GB11, Odin2}.

Recently, a  new dark energy model 
has been proposed \cite{Lim:2010yk,Gao:2010gj}. It consists in considering  two scalars where one 
of them is given by a  Lagrange multiplier.
This multiplier puts a natural  constraint on  the  form on the second
scalar field (in particular on its coupling and self-interacting potential) allowing that  the emerging dark energy behavior  evolves in a dust-matter dominated era, as requested by observations going back in the redshift $z$.
 The interesting feature of this approach is that  the whole system contains a single
dynamical degree of freedom and this fact allows to solve  several  shortcomings related to the fine-tuning of $\Lambda$CDM model, among them the cosmological constant 
problem \cite{LM1}.  
It is important to stress that such a 
Lagrange multiplier technique can be related to the existence of  Noether symmetries and then it is a general approach to reduce dynamical systems and find out exact solutions, as we will discuss below \cite{N}.

The extension of $f(R)$ gravity via the addition of a Lagrange multiplier 
constraint has been  proposed in Ref.\cite{Odin21}. 
Such model  can be considered as a 
new version of modified gravity because  dynamics and cosmological solutions 
are different from the standard version of $f(R)$ gravity without such constraint. This result is clear from a dynamical viewpoint: Lagrange multipliers are anholonomic constraints capable of reducing dynamics \cite{arnold,cimento}.
Furthermore, using the Lagrange multiplier  approach   helps in the formulation 
of covariant renormalizable gravity \cite{O}.

In the present paper, we study the Gauss-Bonnet gravity with Lagrange multiplier 
constraints in view to recover realistic dark energy behaviors. Technically, we are considering a scalar-tensor-higher-order gravity where a Lagrange multiplier is considered. We explicitly show that it is possible to derive  new cosmological solutions in this context.
In particular, we compare the accelerating solutions in string-inspired 
Gauss-Bonnet gravity with and without the Lagrange multiplier term. It is demonstrated that, in the 
version with Lagrange multiplier, one gets large number of new accelerating 
cosmologies, including the phantom cosmologies where the dilaton kinetic term is 
canonical. As a more interesting example, the so-called Little Rip cosmology can be recovered. 

The plan of the paper is the following. In Sect. II, we discuss the scalar Gauss-Bonnet gravity in presence of a Lagrange multiplier. In particular, we consider the  Friedmann-Robertson-Walker (FRW) cosmology and derive exact solutions according to the choice of the scalar field $\phi$. The general scheme of cosmological reconstruction of scalar Gauss-Bonnet gravity with Lagrange multiplier is pursued in Sect. III. Here we work out the whole method starting from the cosmological equations up to the final cosmological solutions. 
 In Sec. IV, we rewrite the cosmological Friedmann equations as an autonomous 
system of first order differential equations  and study its critical points.
Sect. V is devoted to the summary  and discussion of the results. 
The general discussion of the Lagrange multiplier method  in view of the Noether Symmetry Approach is reported in Appendix A.

\section{Scalar Gauss-Bonnet gravity  with Lagrange multiplier}

Let us  study accelerating cosmology in string-inspired scalar
Gauss-Bonnet gravity with Lagrange multiplier. To this end, a suitable 
 action has the following form:
\bea
\label{LagS1}
S &= &\int d^4 x \sqrt{-g} \left\{
\frac{R}{2\kappa^2} - \frac{\omega(\phi)}{2} \partial_\mu \phi
\partial^\mu \phi - V(\phi)-\right.\nonumber\\
&-&\left. \varepsilon(\phi) G
- \lambda \left( \frac{1}{2} \partial_\mu \phi \partial^\mu \phi +
U(\phi) \right) \right\}\, .
\eea

Here $\lambda$ is the Lagrange
multiplier scalar, $\varepsilon$, $\omega$ and $U$ are arbitrary scalar functions  and $G$ is Gauss-Bonnet invariant is
\be
G=R_{\mu\nu\alpha\beta}R^{\mu\nu\alpha\beta}-4R_{\mu\nu}R^{\mu\nu}+R^2.
\ee
The above effective action represents a string-inspired gravity which has been  mainly considered
for exponential potentials and without last term (for a single scalar). The
interpretation of above action is related to specific compactification of
superstring theory in four dimensions where, apart from the dilaton scalar field,  one more scalar (given by the Lagrange
multiplier) is considered.

The gravitational field equations are derived by varying with respect to the metric and assume   the following form:

\bea
\label{LagE} 
&&\frac{1}{2\kappa^2} \left(R_{\mu\nu} - \frac{1}{2}
g_{\mu\nu} R \right) =-\frac{ g_{\mu\nu}}{2} \left\{
       \frac{\omega(\phi)}{2} \partial_\rho \phi \partial^\rho \phi +\right. \nonumber\\
&&\left.+V(\phi) +
 \lambda \left( \frac{1}{2} \partial_\rho \phi
\partial^\rho \phi + U(\phi)
\right) \right\} +\\
&&+ \frac{\omega(\phi) + \lambda}{2} \partial_\mu
\phi \partial_\nu \phi +
4(R_{\mu\nu}-\frac{1}{2}R g_{\mu\nu})\Box\varepsilon(\varphi)+\nonumber\\
&&+
4(R_{\mu\phantom{\alpha\beta}\nu}^{\phantom{\mu}\alpha\beta}-R^{\alpha\beta}
g_{\mu\nu})\nabla_\alpha\nabla_\beta\varepsilon(\varphi)-
8R^\alpha_{\phantom{\alpha}\mu}\nabla_\alpha\nabla_\nu\varepsilon(\varphi)+\nonumber\\
&&+
2R\nabla_\mu\nabla_\nu\varepsilon(\varphi)
\, .\nonumber
\eea

Let us now consider a  FRW metric with
flat spatial part:
\be
\label{Lag7} ds^2 = - dt^2 + a(t)^2 \sum_{i=1,2,3}
\left(dx^i\right)^2\, .
\ee
Then by the variation over $\lambda$,
we obtain the further equation
\be
\label{LagS2} 0 = \frac{{\dot\phi}^2}{2} - U(\phi)\, .
\ee

 The cosmological Friedmann equations are then
\bea
\label{LagS3}
\frac{3}{\kappa^2} H^2 &=& \frac{\omega(\phi) +
\lambda}{2}{\dot\phi}^2 + V(\phi) + \lambda U(\phi)+24 H^3 \frac{d
\varepsilon}{dt} =\nonumber\\
&=& \left(
\omega(\phi) + 2\lambda \right) U(\phi) + V(\phi)+24 H^3 \frac{d
\varepsilon}{dt} \, ,
\eea
\bea
\label{LagS4}
&-& \frac{1}{\kappa^2} \left(2 \dot H + 3H^2 \right)
= \frac{\omega(\phi) + \lambda}{2}{\dot\phi}^2 - V(\phi) -\nonumber\\
&-&\lambda U(\phi) -8 H^2 \frac{d^2 \varepsilon}{dt^2}-16 H \dot{H} \frac{d
\varepsilon}{dt}-16 H^3 \frac{d \varepsilon}{dt}=\\
&=& \omega(\phi) U(\phi) - V(\phi) -8 H^2 \frac{d^2 \varepsilon}{dt^2}-16 H \frac{d \varepsilon}{dt}\left(
\dot{H} + H^2 \right)\, .\nonumber
\eea

It is straightforward to derive, from the above equations,  the effective  EoS-parameter
$w$  as:
\be
w = \frac{ \omega(\phi) U(\phi) - V(\phi) -8 H^2 \frac{d^2
\varepsilon}{dt^2}-16 H\frac{d \varepsilon}{dt}\left(\dot{H} + H^2\right)}{ \left(
\omega(\phi) + 2\lambda \right) U(\phi) + V(\phi)+24 H^3 \frac{d
\varepsilon}{dt} }\, .\nonumber
\ee

The equation obtained by varying the action over the  scalar field  $\phi$ (the Klein-Gordon equation) is a
consequence of second Friedmann   Eq.(\ref{LagS4}) and then is easily achieved.

Eqs.(\ref{LagS3}-\ref{LagS4}) can be easily  solved. Indeed, there are
unknown functions: $H$, $V$, $\phi$, $\lambda$, $\omega$ and $\varepsilon$
which may be used to satisfy two equations.
    Therefore, if we want to obtain a closed system of equations, we must fix
some of the functions  $V$, $\lambda$, $\omega$ and $\varepsilon$ or select the
field
$\phi$ and  metric $a$, as well as two of the functions $V$, $\lambda$,
$\omega$ and $\varepsilon$.
    For example, one can select:
\be
\label{LagS10}
\omega(\phi) = 1\, ,\quad U(\phi) =
\frac{m^4}{2}\, .
\ee

Here $m$ is a constant with the dimension of
mass. Also, the  canonical scalar is considered. Then Eq. (\ref{LagS2})
gives
\be
\label{LagS11}
\phi = m^2 t\, .
\ee

In this case, one can express the potential $V$ and the Lagrange multiplier
$\lambda$  in terms of the Hubble rate $H$ and the function $\varepsilon$:
\bea
V&=&\frac{m^4}{2}+\frac{2\dot{H}}{\kappa^2}
-16 H \dot{\varepsilon}\left( \dot{H} + H^2 \right)
+H^2\left(\frac{3}{\kappa^2}-8\ddot{\varepsilon}\right),\nonumber\\
\lambda&=&-1+\frac{8\kappa^2 H\left(\dot{\varepsilon}(2\dot{H}-H^2)+H
\ddot{\varepsilon}\right)-2\dot{H}}{\kappa^2 m^4}.
\eea
On the other hand, one can select the scalar field potential and the function $\varepsilon$,
following string-inspired considerations:
\be V=\Lambda+V_0 e^{b_1 \phi},\;\;\; \varepsilon=\varepsilon_0 e^{b_2 \phi}.
\label{eqq1}\ee
Here $b_1$ and $b_2$ are  constants.
By analogy, the Lagrange multiplier is chosen in the same form
\be \lambda=\lambda_1+\lambda_0  e^{ b_3 \phi}, \label{eqq2}\ee
with $b_3$ being some constant.

Let us consider now some explicit cases to achieve solutions for the above dynamical system.  It is clear that this is a sort of "inverse scattering procedure", that is by fixing the form of the scalar field, we solve dynamics. Usually, it is exactly the contrary: by fixing scalar field potentials and couplings, field equations gives the scalar field form.

\subsection{The case $\phi \sim t$}

This choice of scalar $\phi$  corresponds to a constant $U(\phi)$. In addition,
we assume that $\omega=\pm 1$. Hence,

\be
\phi = \phi_0 t,\,\,\,\, \omega(\phi)=\gamma=\pm 1 , \label{eqq3}
\ee
where $\phi_0$ is some constant.

One can solve the first Friedmann  equation for the Hubble rate. In this case, the
second equation becomes constant. We get:
\bea
H&=&-\frac{e^{-\phi_0 b_2 t}}{24 \phi_0 \varepsilon_0  b_2 \kappa^2} \left(\frac{1}{(1-2\alpha+\sqrt{4\alpha(\alpha-1)})^{1/3}}+\right.\nonumber\\
&+&\left.(1-2\alpha+\sqrt{4\alpha(\alpha-1)})^{1/3}+1\right),
\label{eqq5}
\eea
here $\alpha=72 \phi_0^2\varepsilon_0^2  b_2^2  \kappa^6  e^{2\phi_0 b_2 t}
\left(\phi_0^2(\gamma+2\lambda)+2V\right)$.

It is easy to see that  $\alpha$  must be either less than zero or greater than
one. That is
$\phi_0^2(\gamma+2\lambda)+2V<0$ or $\phi_0^2(\gamma+2\lambda)+2V>1$.

Choosing the constant case, one can easily meet these requirements. If we assume that
$\alpha$ is much bigger than one, then
\be H \sim e^{- b_2\phi}\sim 1/\varepsilon.\ee
For this case, it is easy to solve the Friedmann equation.
Let the Hubble parameter is given by
\be
\label{eqq6}
H={h_0}e^{\phi_0 h t}. \ee

This is the so-called {\it Little Rip cosmology} \cite{LR11} which can be realized
also in the modified gravity context (see Refs.\cite{AM}).
The effective EoS  parameter of this  model  has the phantom nature, that means
\be w_{eff}=-1-\frac{2\phi_0h}{3h_0}e^{-\phi_0 ht}.
\ee

In the Little Rip cosmology, $w$ approaches $-1$ sufficiently rapidly,
so that it is possible to have a model in which $\rho_{DE}$ increases with
time,
but in which there is no finite-time future singularity.   The characteristic
feature of such a cosmology is  disintegration of bound objects at finite time
(like in the case of {\it Big Rip cosmology}).

As the universe expands, the relative acceleration between two points separated
by a comoving
distance $l$ is given by
$l \ddot a/a$, where $a$ is the scale factor.
An observer sitting at comoving distance $l$ away from a mass $m$ will measure
an inertial force on the mass of the order
\be
\label{i1}
F_\mathrm{iner}=m l \ddot a/a = m l \left( \dot H + H^2 \right)\, .
\ee
Let us assume the two point force is bounded by the bounding force $F_0$. If
$F_\mathrm{iner}$ is positive and greater
than $F_0$, the two particles become unbound. This effect  is the ``Rip''
produced by the accelerating expansion.

Eq. (\ref{i1}) shows that  Rip occurs when either $H$ diverges or
$\dot H$
diverges (assuming $\dot H > 0$) \cite{LR11,Frampton:2011rh}.
As we see,  it is possible for $H$
and for
$F_\mathrm{iner}$, to increase without bound and yet not produce a future
singularity at a finite time. This is the essence of Little Rip.

From Eq.~(\ref{i1}) for (\ref{eqq6}), we have
\be
\label{i5}
F_\mathrm{iner}=m l \left( h_0 \phi_0\, h\, \e^{ \phi_0\, h\, t} + h_0^2 \e^{2
\phi_0\, h \,t} \right)\, ,
\ee
which is positive and unbounded. Thus, $F_\mathrm{iner}$
becomes arbitrary large with increasing $t$, that results in a Little Rip.
The parameter  $\phi_0 h$ can be estimated   \cite{LR11}. Let us assume
\be \phi_0 h=\frac{\sqrt{3} \beta}{2},\,\,\,\hbox{where} \,\, \beta\sim 3.46
\times 10^{-3} Gyr^{-1}\,.\ee
In addition, the present value of the Hubble constant is $H_0^{-1}=13.6\;
Gyr^{-1}$.
In this case, one can estimate the time required for the disintegration of an
object of the size of the Solar System as $t\sim 7750 \; Gyr$.
Furthermore, it is easy to get the theory constants values $\phi_0 h=2.99
\times 10^{-3}$, $h_0=70.59\times 10^{-3}$.
For such a set of constants, the EoS parameter is equal to $-1.0272$ and this value  does
not contradict the observational bounds.

Let us make some general comments about the impact of the Lagrange
multipliers on $F_{iner}$. From (\ref{i1}) and (\ref{LagS3}) one  finds
\bea \label{eqq15}
&&H^2+\dot{H} =\\\
&& \kappa^2\left(-\frac{2}{3}\gamma U-\frac{1}{3} \lambda
U+\frac{1}{3} V+4 H^3 \dot{\varepsilon}+8H\dot{\varepsilon}\dot{H}+4H^2
\ddot{\varepsilon}\right).\nonumber
\eea

It can be seen that, depending on  the sign,  $\lambda$  can either enhance or
decrease $F_{iner}$. In other words, for realistic models, the role of Lagrange
multiplier may be related to the increase/decrease of the time left before the
disintegration of bound objects in the universe.

Let us now find the solution of the Friedmann equations (\ref{LagS3}-\ref{LagS4}) for the
model (\ref{eqq6}) asuming the exponential scalar functions (\ref{eqq1}-\ref{eqq2}).
There are two solutions:
\bea
\label{eqq10}
V_1&=&\frac{1}{2}\phi_0^2\gamma-\frac{H^2}{\kappa^2}\;\;\quad\mbox{or}\quad\;\;
V_2=\frac{1}{2}\phi_0^2\gamma+3\frac{H^2}{\kappa^2},\nonumber\\
\varepsilon_1&=&-\frac{1}{4 \phi_0 h\;  h_0 \kappa^2}\frac{1}{H}\;\;\quad\mbox{or}\quad\;\;
\varepsilon_2=-\frac{1}{16 \kappa^2}\frac{1}{H^2},\\
\lambda_1&=&-\gamma-\frac{2}{\phi_0^2 \kappa^2} H^2\;\;\quad\mbox{or}\quad\;\;
\lambda_2=-\gamma-\frac{3h}{\phi_0 \kappa^2} H.\nonumber
\eea

In the first case,  $\alpha$  tends to a constant value, which corresponds to
the solution (\ref{eqq5}). In the second case $\alpha\to 0$.

Now we consider the case when $\varepsilon=0$.  This solution is studied
in Ref. \cite{Odin2}. Selecting the potential and Lagrange multiplier as
(\ref{eqq1}) and (\ref{eqq2}), we do not get the solution in the form $H \sim
e^\phi$.
However, it is easy to choose the potential in a slightly different form to get
again  the exponential Hubble rate.

Indeed, we have
\bea \label{eqq11}
H&=&h_0 e^{\phi_0 h t},\nonumber\\
V&=&\frac{1}{2}\phi_0^2\gamma+\frac{2\phi_0 h h_0}{\kappa^2} e^{\phi_0 h
t}+\frac{3 h_0^2}{\kappa^2} e^{2\phi_0 h t},\\
\varepsilon&=&0,\nonumber\\
\lambda&=&-\gamma-\frac{2 h_0 h}{\phi_0 \kappa^2} e^{ \phi_0 h t}.\nonumber
\eea

Once again, the Friedmann equations are satisfied. Obviously, this scheme may be applied to
generate new cosmological solutions. It is clear that the potential and the
Lagrange
multiplier have now the form
\bea
\label{LagS12}
V &=& \frac{1}{\kappa^2}\left\{ 2 \dot H(t) + 3 H(t)^2 \right\} +
\frac{\phi_0^2\gamma}{2}\, ,\nonumber\\
\label{LagS13} \lambda &=& -1 - \frac{2}{\kappa^2} \dot H(t) \, .
\eea
This solution has been obtained in ref.\cite{Odin1} in  the same
way.

As an example one can keep just a Lagrange multiplier ($V=0$ and
$\varepsilon=0$) in the form  (\ref{eqq2}). Then solving the Friedmann equations, it is
easy to find the Hubble rate
\be
H=-\frac{\phi_0\sqrt{\gamma} \kappa}{\sqrt{6}}
\tan{\left[\frac{1}{2}\sqrt{\frac{3}{2}} \phi_0 \sqrt{\gamma} \kappa (t -
const)\right] }.
\ee

Since the solution is derived from the second Friedmann Eq. (\ref{LagS4}), there
is no dependence on the Lagrange multiplier. Now using a Lagrangian multiplier,
one can convert the first Friedmann Eq. (\ref{LagS3}) into an identity.
In this case, we obtain $\lambda$ as follows
\be
\lambda=\frac{1}{2}\gamma\left(-1 +\tan{\left[\frac{1}{2}\sqrt{\frac{3}{2}}
\phi_0 \sqrt{\gamma} \kappa (t - const)\right]}^2\right).
\ee
For this solution, EoS parameter $w$ lies in the range of $-1/3$ to infinity.
In other words, the expansion is decelerating.

There is another solution, if $\gamma = -1$ (in this case, $H=$const).  The
presence of the Lagrange multiplier leads to new solutions which do not exist
without it. For example, using the conditions (\ref{eqq1}), (\ref{eqq3}) and
(\ref{eqq6}) one cannot get a solution without a Lagrangian multiplier. If
Lagrange multiplier
    (\ref{eqq2}) is taken into account then it is easy to construct the solution
(\ref{eqq10}) and (\ref{eqq11}). This means that the presence of the Lagrange multiplier effectively changes the dynamical system giving rise to a constrained dynamics with, in principle,  different solutions.

\subsection{The case $\phi \sim \ln{t}$}

This form of scalar field allows us to choose the Hubble rate as $1 / t$ for the
functions $V$, $\varepsilon$ and $\lambda$ given by Eqs. (\ref{eqq1})  and (\ref{eqq2}).

That is
\be H=\frac{h_0}{t},\;\;\; \phi=\phi_0 \ln{\frac{t}{t_1}},\,\,\,\,
\omega(\phi)=\gamma=\pm 1 , \label{rec1}
\ee
where $h_0$, $\phi_0$ and $t_1$ are some constants. When $h_0 >0$, we have a
 quintessential power-law expansion. For
\be H=-\frac{h_0}{t_s-t},\;\;\; \phi=\phi_0 \ln{\frac{t_s-t}{t_1}},\,\,\,\,
\omega(\phi)=\gamma=\pm 1 , \label{BRT}
\ee
we have a phantom power-law expansion when $h_0<0 $.
A phantom model describes the Big Rip finite-time future singularity:
for $t =t_s$ the scale factor  tends to infinity.

Now we can write the Friedmann equations as 
\bea
&-&\Lambda-\frac{\phi_0^2 \gamma}{2 t^2}+\frac{3 h_0^2}{k^2 t^2}-\frac{\phi_0^2
\lambda_1}{t^2}-24 \varepsilon_0  b_2 h_0^3 t^{-4+ b_2} t_1^{- b_2}-\nonumber\\
&-&\phi_0^2
\lambda_0 t^{-2+ b_3} t_1^{- b_3}-t^{b_1} t_1^{-b_1} V_0=0,\nonumber\\
\label{eeqq1}
&-&\Lambda+\frac{\phi_0^2 \gamma}{2 t^2}-\frac{2 h_0}{k^2 t^2}+\frac{3 h_0^2}{k^2
t^2}+\\
&+&8 \varepsilon_0  b_2 h_0^2\left(3-  b_2 -2 h_0 \right) t^{-4+ b_2} t_1^{-
b_2}-t^{b_1} t_1^{-b_1} V_0=0.\nonumber
\eea

For $h_0>0$  (for $h_0<0$ we have to replace $t$ with $t_s-t$)   and
\bea V&=&\Lambda+V_0 e^{b_1 \phi/\phi_0},\;\;\; \varepsilon=\varepsilon_0 e^{ b_2
\phi/\phi_0},\nn \lambda&=&\lambda_1+\lambda_0  e^{ b_3 \phi/\phi_0}
\label{eq3}\eea

First, we consider the case when the Lagrangian multiplier and the potential is
zero ($\lambda=V=0$). It is easy to see, that
\bea &&b_2=2,\,\,\,\phi_0^2=-\frac{6h_0^2(1-h_0)}{\gamma\kappa^2(1-5h_0)}\\
&&\varepsilon=\frac{(3h_0-1)}{8h_0(5h_0-1)\kappa^2}\left(ts-t\right)^2,\;
h_0<0 \nonumber\\
&& \mbox{or}\nn
&&\varepsilon=\frac{(3h_0-1)}{8h_0(5h_0-1)\kappa^2}\left(t\right)^2,
\;h_0>0\label{wlm1}.\nonumber
\eea
The solutions exist if

a) $\gamma$ <0 and $h_0$<0 or 0<$h_0$<1/5 (or $h_0$>1),

b) $\gamma$ >0  and 1/5<$h_0$ <1.

In the case in which the scalar field $\phi$ is canonical ($\gamma$=1), and there is no
potential ($V(\phi)$=0), we cannot obtain the effective phantom cosmological
solution with $w<-1$. A similar situation is obtained  for $\varepsilon=0$ and
$V\ne 0$. In this case, the potential has the form
\bea V&=&\frac{h_0(3h_0-1)}{\kappa^2 t^2},\; \mbox{for}
\;\;h_0>0\nn
\mbox{or}\;\;V&=&\frac{h_0(3h_0-1)}{\kappa^2 (t_s-t)^2},\; \mbox{for}\;\; h_0<0.
\label{wlm2}\eea
In addition there are restrictions on $h_0$: $\phi_0^2 \gamma=2h_0/\kappa^2$.
One sees that $w<-1$ only in the case $\gamma<0$.

Now let $\lambda=0$ but $V\ne 0$ and $\varepsilon\ne 0$. In this case, the
solution exists only for $ b_2=-b_1=2$  \cite{Odin1}. Then
\bea
\varepsilon&=&\frac{\left(2 h_0-\phi_0^2 \gamma k^2\right) t^2}{16 h_0^2
(1+h_0) k^2},\nonumber\\
V&=&\frac{6 (-1+h_0) h_0^2+\phi_0^2 \gamma (-1+5 h_0) k^2}{2 (1+h_0) k^2 t^2},\\
&&h_0>0\;\;\mbox{or}\;\; t\to t_s-t,\;\;\mbox{if} \;\;h_0<0. \label{wlm3}\nonumber
\eea
It is evident that there is a phantom solution for $\gamma$> 0.

If $\lambda\ne 0$  and $V=0$,  we obtain the following conditions restricting
$h_0$
\be \phi_0^2=-\frac{2h_0^2(3h_0-2)}{\gamma\kappa^2}.
\ee

Therefore, to get the value of $w$  about $-1$, $\gamma$ must be negative. If
$\gamma>0$, then $h_0$ has to lie in the following range $0<h_0<2/3$.
For this case, we have
\bea
\varepsilon&=&\varepsilon_0\left(\frac{t}{t_1}\right)^{3-2h_0},\nonumber\\
\lambda&=&\frac{\gamma(1-3h_0)}{3h_0-2}-\frac{12\varepsilon_0\gamma
h_0^2(2h_0-3)}{(3h_0-2)t_1^2}\left(\frac{t}{t_1}\right)^{1-2h_0}\\
&&\mbox{for}\;\;
h_0>0 \;\; \mbox{and}\;\; t\to t_s-t\;\; \mbox{if}\;\; h_0<0 \label{lm3}.\nonumber
\eea
For sufficiently large $h_0$, the function $\varepsilon$ has a negative power and
quickly decreases with time (for$ h_0> 0$) or increases when approaching to
$t_s$ (for $h_0 <0$). For the model without a Lagrangian multiplier
(\ref{wlm2}),
function $\varepsilon$ grows for $h_0> 0$ and decreases for $h_0 <0$.

If $\lambda\ne 0$  and $\varepsilon=0$,  we get
\be
\lambda=-\gamma+\frac{2h_0}{\phi_0^2\kappa^2},\;\;V=\frac{-4h_0+6h_0^2+\phi_0^2
\gamma\kappa^2}{2\kappa^2 t^2}. \label{lm4} \ee
Such behavior is similar to the one in the model without the Lagrange
multiplier (\ref{wlm1}).

Suppose now that the action has the form (\ref{LagS1}), that is $V\ne 0$,
$\varepsilon\ne 0$ and $\lambda\ne 0$.
    From Eqs.(\ref{eeqq1}), one sees that there are several solutions:

1) $b_1=-4+ b_2$. For this case, there are two additional conditions: $ b_2=2$
and $ b_2 \ne 2$.

In the first case it is easy to find:
\bea
V&=&\frac{32 \varepsilon_0
(1-2h_0)h_0^2\kappa^2+(2h_0(3h_0-2)+\phi_0^2\gamma\kappa^2)t_1^2}{2\kappa^2
t_1^2 t^2},\nonumber\\
\varepsilon&=&\varepsilon_0\frac{t^2}{t_1^2}, \label{lm5} \\
\lambda&=& \frac{-16\varepsilon_0
h_0^3(1+h_0)\kappa^2+(2h_0-\phi_0^2\gamma\kappa^2)t_1^2}{\phi_0^2 \kappa^2
t_1^2}=const.\nonumber\eea

This means that  the solution is similar to the solution for the case without
Lagrange multiplier (\ref{wlm3}). The time dependence is the same.
The difference enters via  an arbitrary factor $\varepsilon_0$  by choosing the
Lagrange multiplier.

Now, let $ b_2 \ne 2$, then we obtain the following condition
\be \phi_0=\pm \frac{\sqrt{2}}{\kappa}\sqrt{\frac{h_0(2-3h_0)}{\gamma}}.\ee

Again  limits appear on the Hubble rate. Phantom cosmology is possible
only for $\gamma<0$. If  $\gamma>0$,
then $h_0$ must be greater than zero and less than $2/3$. For this case, we
have
\bea
V&=&-\frac{8 \varepsilon_0  b_2 h_0( b_2-3+2h_0)(t/t_1)^{-4+
b_2}}{t_1^{4}},\nonumber\\
\varepsilon&=&\varepsilon_0\frac{t^{ b_2}}{t_1^{ b_2}},\label{lm6}\\
\lambda&=&\gamma\frac{1-3h_0}{3h_0-2}+\frac{4\varepsilon_0 b_2\gamma(
b_2-3-h_0)h_0\kappa^2}{(2-3h_0)t_1^2}\left(\frac{t}{t_1}\right)^{-2+
b_2}.\nonumber\eea

2) Assume now  the following conditions $ b_2=3-2h_0$. Then
\bea
V&=&\frac{-4h_0+6 h_0^2+\phi_0^2 \gamma \kappa^2}{2 \kappa^2 t^2},\nonumber\\
\varepsilon&=&\varepsilon_0 \left(\frac{t}{t_1}\right)^{3-2 h_0},\label{lm7}\\
\lambda&=&-\gamma+\frac{2 h_0}{\phi_0^2 k^2}+\frac{24 \varepsilon_0 h_0^3 (-3+2
h_0) \left(\frac{t}{t_1}\right)^{1-2 h_0}}{\phi_0^2 t_1^2}.\nonumber\eea

3)  There is one more case: $ b_2=4$. Then
\bea
V&=&-\frac{32\varepsilon_0 h_0^2(1+2h_0)}{t_1^4}+\frac{-4h_0+6 h_0^2+\phi_0^2
\gamma \kappa^2}{2 \kappa^2 t^2},\nonumber\\
\varepsilon&=&\varepsilon_0 \left(\frac{t}{t_1}\right)^{4},\label{lm8}\\
\lambda&=&-\gamma+\frac{2 h_0}{\phi_0^2 k^2}+\frac{32 \varepsilon_0 h_0^2 (1-
h_0) t^2}{\phi_0^2 t_1^4}\nonumber.\eea

Hence, the Lagrange multiplier presence  helps to generate new cosmological
solutions. For models
without such multipliers,
    we find that the potential is proportional to the square of the time, while
the scalar function of the Gauss-Bonnet invariant  is inversely proportional to
the time. In the model with the Lagrange multiplier we get new solutions
(\ref{lm3}), (\ref{lm6}), (\ref{lm7}) and (\ref{lm8}).  In all these cases, the
dependence of the coefficient of the Gauss-Bonnet invariant is explicit and
simple, whereas the scalar potential often shows the same time dependence
($V\sim 1/t^2$).
It should also be noted that there may occur phantom cosmologies for the
canonical scalar field $\phi$. Only for solutions (\ref{lm6}) we obtain
restrictions on $h_0$ (for
$V$, $\varepsilon$ and $\lambda$ $\ne 0$).

Without the Lagrange multiplier, the scalar potential increases with time as $t^2$ (if
$h_0> 0$), and the function $\varepsilon$ decreases as $1 /t^2$, or vice-versa
if $h_0 <0$. In  presence of the Lagrange multiplier, one can get  exactly
the same behavior as well as new solutions:
    $\varepsilon$ and $V$   increase (for $h_0> 0$) or another case where the
potential decreases as $1 /t^2$, and $\varepsilon$ increases as $t^{3-2h_0}$
(for $h_0 <0$) or decreases (for $h_0 <0$).
Let us consider the case with/without the Lagrange multiplier but with the same
scalar potentials. Let $b_1=-2,\;\;b_2=2$ and
    $h_0=-80/3,\;\kappa=1,\;\gamma=1$.
In this case, we numerically derive 
\bea
V_0&=&2304.8 +4.23377\, \lambda\nonumber\\
\varepsilon&=&
0.000189478 + 6.84862*10^{-6} \lambda.
\eea
The Big Rip cosmology occurs with the corresponding Rip time for model
(\ref{BRT}) at
$t_s=376.27$ Gyr.
In summary, the presence of Lagrange multiplier may lead to the generation of new cosmological solutions if compared with scalar Gauss-Bonnet
gravity without such term. The general reason for this different behavior will be discussed below.

\section{The reconstruction of scalar-Gauss-Bonnet gravity with Lagrange
multiplier}
In this section we discuss the  general scheme of cosmological reconstruction in
the scalar-Gauss-Bonnet gravity with the Lagrange multiplier. Let $\omega(\phi)$ be 1.
Friedmann  Eqs. (\ref{LagS3}-\ref{LagS4}) can be written as
\bea
\label{1LagS3}
&&\frac{3}{\kappa^2} H^2=\nn
&&= \left(
1 + 2\lambda(\phi) \right) \frac{\dot{\phi}^2}{2} + V(\phi)+24 H^3 \frac{d
\varepsilon(\phi(t)) }{dt} \, ,\\
\label{1LagS4}
&&\frac{-1}{\kappa^2} \left(2 \dot H + 3H^2 \right)=    \frac{\dot{\phi}^2}{2}
- V(\phi) -8 H^2 \frac{d^2 \varepsilon(\phi(t)) }{dt^2}-\nn
&&-16 H \dot{H} \frac{d
\varepsilon(\phi(t)) }{dt}-16 H^3 \frac{d \varepsilon(\phi(t)) }{dt}\, .
\eea

It is not difficult to express the Lagrange multiplier and the potential in terms of the other
functions.
In this way we get them by fixing the metric, the scalar field and the
coefficient of the Gauss-Bonnet invariant. They are
\bea
\label{2LagS3}
&&\lambda =\\
&&=\dot{\phi}^{-2}\left(- \frac{2}{\kappa^2} \dot H-
\frac{\dot{\phi}^2}{2}- 8 H^2 \frac{d^2 \varepsilon}{dt^2}-8 H  \frac{d
\varepsilon}{dt}(\dot{H}-H^2) \right)\, ,\nn
\label{2LagS4}&&V
=  \\
&&=\frac{2 \dot H + 3H^2 }{\kappa^2} + \frac{\dot{\phi}^2}{2}
-8 H^2 \frac{d^2 \varepsilon }{dt^2}-16 H  \frac{d \varepsilon }{dt}(\dot{H}+H^2)\, .\nonumber
\eea

Using these expressions, it is easy to verify the validity of earlier solutions.

Let us consider now the case when, as  unknown functions do not appears  the  Lagrange
multiplier and the function $\varepsilon(\phi)$.
Combining the Friedmann Eqs.  (\ref{1LagS3}) and (\ref{1LagS4}), we obtain
\bea
\label{GBany7}
0&=&\frac{2}{\kappa^2}\dot H + {\dot\phi}^2 +\lambda(\phi)\dot{\phi}^2 - 8H^2
\frac{d^2
\varepsilon(\phi(t))}{dt^2}-\nn
     &  - &16 H\dot H \frac{d\varepsilon(\phi(t))}{dt} + 8H^3
\frac{d\varepsilon(\phi(t))}{dt}\\
&=&\frac{2}{\kappa^2}\dot H + {\dot\phi}^2 +\lambda(\phi)\dot{\phi}^2 -
8a\frac{d}{dt}\left(\frac{H^2}{a}\frac{d\varepsilon(\phi(t))}{dt}\right)\
.\nonumber
\eea
Eq.(\ref{GBany7}) can be solved with respect to $\varepsilon(\phi(t))$ as
\bea
\label{GBany8}
&&\varepsilon(\phi(t))=\frac{1}{8}\int^t dt_1 \frac{a(t_1)}{H(t_1)^2} \times\\
&&\times\int^{t_1}
\frac{dt_2}{a(t_2)}
\left(\frac{2}{\kappa^2}\dot H (t_2) + {\dot\phi(t_2)}^2
+\lambda(\phi(t_2))\dot{\phi}(t_2)^2\right)\ .\nonumber
\eea
Combining Eqs. (\ref{1LagS3}) and (\ref{GBany8}), the scalar potential $V(\phi(t))$
is:
\bea
\label{GBany9}
&&V(\phi(t)) = \frac{3}{\kappa^2}H(t)^2 - \frac{1}{2}{\dot\phi (t)}^2
-\lambda(\phi (t)){\dot\phi (t)}^2- \nn
&&-3a(t)
H(t) \int^t \frac{dt_1}{a(t_1)}\times\\
&&\times
\left(\frac{2}{\kappa^2}\dot H (t_1) + {\dot\phi(t_1)}^2
+\lambda(\phi(t_1))\dot{\phi}(t_1)^2\right)\ .\nonumber
\eea

Let us identify $t$ with $f(\phi)$ and $H$ with $g'(t)$ where $f$ and $g$
are some unknown functions.  
The solution of the Friedmann equations is  related to the  existence and behavior  of
such functions.
Then we consider a model where $V(\phi)$ and $\varepsilon(\phi)$ can  be
expressed
in terms of the two functions $f$ and $g$ as
\bea
\label{GBany10b}
V(\phi) &=& \frac{3}{\kappa^2}g'\left(f(\phi)\right)^2 -
\frac{1}{2f'(\phi)^2}-\frac{\lambda(\phi)}{f'(\phi)^2}-\nn
& - &3g'\left(f(\phi)\right) \e^{g\left(f(\phi)\right)} \int^\phi d\phi_1
f'(\phi_1 ) \e^{-g\left(f(\phi_1)\right)} \times\nonumber\\
& \times&\left(\frac{2}{\kappa^2}g''\left(f(\phi_1)\right) + \frac{1}{f'(\phi_1
)^2}+\frac{\lambda(\phi_1)}{f'(\phi_1 )^2}
\right)=\nn
&=&\frac{3}{\kappa^2}g'\left(f(\phi)\right)^2 -
\frac{1}{2f'(\phi)^2}-\nn&- &3g'\left(f(\phi)\right) \e^{g\left(f(\phi)\right)} \int^\phi d\phi_1
f'(\phi_1 ) \e^{-g\left(f(\phi_1)\right)} \times\nonumber\\
& \times&\left(\frac{2}{\kappa^2}g''\left(f(\phi_1)\right) + \frac{1}{f'(\phi_1
)^2}\right)-\nn
&-&\frac{\lambda(\phi)}{f'(\phi)^2}- 3g'\left(f(\phi)\right)
\e^{g\left(f(\phi)\right)}\times\nn
&\times& \int^\phi d\phi_1
\e^{-g\left(f(\phi_1)\right)} \left(\frac{\lambda(\phi_1)}{f'(\phi_1 )}
\right)\ ,
\eea
\bea
\varepsilon(\phi) &=& \frac{1}{8}\int^\phi d\phi_1 \frac{f'(\phi_1)
\e^{g\left(f(\phi_1)\right)} }{g'(\phi_1)^2}\times\nn
&\times&
\int^{\phi_1} d\phi_2  f'(\phi_2) \e^{-g\left(f(\phi_2)\right)}\times\nn
&\times&
\left(\frac{2}{\kappa^2}g''\left(f(\phi_2)\right) +
\frac{1}{f'(\phi_2)^2}+\frac{\lambda(\phi_2)}{f'(\phi_2 )^2}
\right)=\nn
&=& \frac{1}{8}\int^\phi d\phi_1 \frac{f'(\phi_1)
\e^{g\left(f(\phi_1)\right)} }{g'(\phi_1)^2}\times\nn
&\times&\int^{\phi_1} d\phi_2  f'(\phi_2) \e^{-g\left(f(\phi_2)\right)}\times\nn
&\times&\left(\frac{2}{\kappa^2}g''\left(f(\phi_2)\right) +
\frac{1}{f'(\phi_2)^2}\right)+\nn
&+& \frac{1}{8}\int^\phi d\phi_1 \frac{f'(\phi_1)
\e^{g\left(f(\phi_1)\right)} }{g'(\phi_1)^2}\times\nn&\times&
\int^{\phi_1} d\phi_2   \e^{-g\left(f(\phi_2)\right)}
\left(\frac{\lambda(\phi_2)}{f'(\phi_2 )}
\right)\ . \label{eeee1}
\eea
By choosing $V(\phi)$ and $\varepsilon(\phi)$ as in Eqs. (\ref{GBany10b}), one can
easily find
the following solution for Eqs.(\ref{1LagS3}) and (\ref{1LagS4}) that can be compared with results in
\cite{Odin2}), that is
\bea
\label{GBany11b}
\phi=f^{-1}(t)\quad \left(t=f(\phi)\right)\ ,\nn
a=a_0\e^{g(t)}\ \left(H= g'(t)\right)\ .
\eea

Hence  any monotonic evolving cosmology,  expressed by $H=g'(\phi)$ in the model (\ref{LagS1}) with the potential
(\ref{GBany10b}), can be realized, including  models exhibiting the transition from non-phantom phase to phantom
phase without introducing the phantom scalar field. However, we have to note that the approach could become problematic for $H(t)$ non-monotonic as we will discuss in the next subsection.

As an example, let us  consider the model (\ref{rec1}). Hence,
\bea
f(\phi)&=&t=t_1 e^{\phi/\phi_0},\;\;f'(\phi)=\frac{t_1
}{\phi_0}e^{\phi/\phi_0},\nn
g(\phi)&=&h_0\frac{\phi_0}{\phi},\;\;\;g'(\phi)=\frac{h_0}{t_1}e^{-\phi/\phi_0},
\eea
where prime is the time derivative ($'=d/dt$). Now we choose the Lagrange
multiplier as in Eqs. (\ref{lm8}). In this case, the potential and the function
$\varepsilon$ iare composed by two terms: the first term is the value in the
absence of a Lagrange multiplier.
   For   $\varepsilon$, it  has the form
\be
\frac{t_1^2 (2h_0-\phi_0^2 \kappa^2)}{16
h_0^2(1+h_0^2)\kappa^2}e^{2\phi/\phi_0}.\nonumber
\ee
If  $e^{2\phi/\phi_0}=t^2/t_1^2$, this expression is equal to (\ref{wlm3}).
The second term (\ref{eeee1}) has the form
\be
e^{2\phi/\phi_0}\frac{\left(16 e^{2\phi/\phi_0} \varepsilon_0
h_0^2(1+h_0)\kappa^2+(\phi_0^2
\kappa^2-2h_0)t_1^2\right)}{16h_0^2(1+h_0)\kappa^2}.\nonumber
\ee
It also contains the first term with a different sign, and also the following
expression
\be \varepsilon_0 e^{4\phi/\phi_0}= \varepsilon_0 \left(\frac{t}{t_1}\right)^4,
\ee
which coincides with Eq.(\ref{lm8}).
This example is paradigmatic to show how the reconstruction scheme works for the Gauss-Bonnet gravity with
Lagrange multiplier.

\subsection{The case of non-monotonic  functions of time.}

The above examples  correspond to  monotonic behaviors in time for the Hubble parameter and the scalar field. 
However, from   Eqs. (\ref{GBany10b}) and (\ref{eeee1}),  one could run into  singularities emerging from non-monotonic behaviors  (for example $\dot{H}=0$ and $\dot{\phi}=0$ for some values of $t$). If we consider the original Eqs.  (\ref{1LagS3}) and (\ref{1LagS4}), we see that terms like $1/\dot{H}$ and $1/\dot{\phi}$ do not appear. The problem may occur when we recast $V$ and $\varepsilon$ in terms of  $\phi$.  To face the issue of non-monotonic behaviors and construct suitable examples, let us take into account the following form of the Hubble parameter
\be
\label{NM_beh1}
H=h_0\left(\frac{1}{t}+\frac{1}{t_0-t}\right).
\ee
This means that the universe is in a non-phantom phase for $t<t_0/2$ and in a phantom phase for  $t>t_0/2$. There is also a Big Rip singularity for $t=t_0$ and a point  where one can define an  effective cosmological constant for $t=t_0/2$ ($\dot{H}(t_0/2)=0$). 

One can take into account  two different behaviors  for the  scalar field:

1) $\phi=\phi_0 t$, that is $t=f(\phi)=\phi/\phi_0$ and

2) $\phi=\phi_0 \ln \left(t(t_0-t)\right)$, that is $t=f(\phi)=$

      $=\frac{1}{2} \left(t_0\pm\sqrt{t_0^2-4 e^{\phi/\phi_0}}\right) $.  Clearly, expressing $V$ and $\varepsilon$ by $\phi$,   uncertainties on the time evolution  can come out while they are removed as soon as   time-dependence in considered.

Let us derive   $V$ and $\varepsilon$ for the first case. The general case contains special functions under the integrals and then  we consider the specific case $h_0 = 2$. We have 
\bea
\varepsilon&=&\frac{1}{32\phi_0^2\phi_s^2}\left(
\frac{\phi^6}{6}+\frac{ c_1 \phi^5}{5}+c_2+\frac{2 \phi^5\phi_s}{25}-\frac{\phi^4 \phi_s^2}{4}+\right.\nonumber\\
&+&\left. \frac{8 \phi^5}{15 \phi_s \kappa^2}-\frac{4 \phi^3\phi_s}{3 \kappa^2}+\frac{2 \phi^2 \phi_s^2}{3 \kappa^2}-\frac{2}{5} \phi^5\phi_s \ln\frac{\phi}{\phi_s}\right)+\nonumber\\
&+& \frac{1}{8}\int^\phi d\phi_1 \frac{f'(\phi_1)
\e^{g\left(f(\phi_1)\right)} }{g'(\phi_1)^2}\times\nn&\times&
\int^{\phi_1} d\phi_2   \e^{-g\left(f(\phi_2)\right)}
\left(\frac{\lambda(\phi_2)}{f'(\phi_2 )}
\right),\\
V&=&
\frac{4 {\phi_0}^2 (2 {\phi}+{\phi_s})^2}{ {\phi}^2 ({\phi}-{\phi_s})^2 {\kappa}^2}-\frac{12 {\phi_0}^2 {\phi} {\phi_s}^2  \ln \frac{\phi}{{\phi_s}}}{ ({\phi}-{\phi_s})^3}+\nonumber\\
&+&\frac{2 {c_1} ({\phi}-{\phi_s})^2+{\phi_0}^2 \left(-{\phi}^2+14 {\phi} {\phi_s}+11 {\phi_s}^2\right) }{2 ({\phi}-{\phi_s})^2 }-\nonumber\\
&-&\frac{\lambda(\phi)}{f'(\phi)^2}- 3g'\left(f(\phi)\right)
\e^{g\left(f(\phi)\right)}\times\nn
&\times& \int^\phi d\phi_1
\e^{-g\left(f(\phi_1)\right)} \left(\frac{\lambda(\phi_1)}{f'(\phi_1 )}
\right)\ ,
\eea
where  $\phi_s=\phi_0 t_0$. We see that by choosing different types of functions $\lambda$, we obtain different forms of  $V$ and $\varepsilon$. It is not always easy to find the resulting integral:  for example, choosing $\lambda =\lambda_0 e^{\phi/\phi_0}$, we obtain the solution in terms of the  Euler function. However, it is easy to obtain solutions for the case
$\lambda =\lambda_0\left(\phi/\phi_0\right)^{\lambda_1}$. We have

\bea
\varepsilon&=&\frac{1}{32\phi_0^2\phi_s^2}\left(
\frac{\phi^6}{6}+\frac{ c_1 \phi^5}{5}+c_2+\frac{2 \phi^5\phi_s}{25}-\frac{\phi^4 \phi_s^2}{4}+\right.\nonumber\\
&+& \frac{8 \phi^5}{15 \phi_s \kappa^2}-\frac{4 \phi^3\phi_s}{3 \kappa^2}+\frac{2 \phi^2 \phi_s^2}{3 \kappa^2}-\frac{2}{5} \phi^5\phi_s \ln\frac{\phi}{\phi_s}+\nonumber\\
&+& \frac{\phi^4 \left(\frac{\phi}{\phi_0}\right)^{\lambda_1}\lambda_0 }{ (\lambda_1-1) \lambda_1 (1+\lambda_1) (4+\lambda_1) (5+\lambda_1) (6+\lambda_1)}\times\nonumber\\
&\times&\left(\phi^2 (\lambda_1-1) \lambda_1 (4+\lambda_1) (5+\lambda_1)-\right.\nonumber\\
&-&2\phi\phi_s (\lambda_1-1) (1+\lambda_1) (4+\lambda_1) (6+\lambda_1)+\nonumber\\
&+&\left.\left.
\phi_s^2 \lambda_1 (1+\lambda_1) (5+\lambda_1) (6+\lambda_1)\right)\right),
\eea
\bea
V&=&
\frac{4 {\phi_0}^2 (2 {\phi}+{\phi_s})^2}{ {\phi}^2 ({\phi}-{\phi_s})^2 {\kappa}^2}-\frac{12 {\phi_0}^2 {\phi} {\phi_s}^2  \ln \frac{\phi}{{\phi_s}}}{ ({\phi}-{\phi_s})^3}+\nonumber\\
&+&\frac{2 {c_1} ({\phi}-{\phi_s})^2+{\phi_0}^2 \left(-{\phi}^2+14 {\phi} {\phi_s}+11 {\phi_s}^2\right) }{2 ({\phi}-{\phi_s})^2 }-\nonumber\\
&-&\frac{\phi_0^2 \left(\frac{\phi}{\phi_0}\right)^{\lambda_1}\lambda_0}{(\phi-\phi_s)^3 \lambda_1 \left(\lambda_1^2-1\right)}\times\nonumber\\
&\times&
 \left(-3 \phi^2 \phi_s (-1+\lambda_1) \lambda_1 (3+\lambda_1)+\right.\nonumber\\
&+&3 \phi \phi_s^2 (-1+\lambda_1) (1+\lambda_1) (4+\lambda_1)-\nonumber\\
&-&\left. \phi_s^3 \lambda_1 (1+\lambda_1) (5+\lambda_1)+\phi^3 \lambda_1 \left(-1+\lambda_1^2\right)\right)\,.
\eea
This means that the delicate point is related to a suitable choice of $\lambda$.

Let us  consider now the second case where the scalar field is a logarithmic function of time. We assume again  $h_0 = 2$ and $\lambda=\lambda_0 e^{-2\phi/\phi_0}$. For this choice, we obtain solutions that do not contain special functions, that is

\bea
\varepsilon&=&\frac{\phi_0^2}{80 t_0^2}\left(
\frac{20c_1 t^7}{7 \phi_0^2}-\frac{5}{4} \phi_0^2 \kappa^2 t^8-\frac{2 t^5 t_0}{\phi_0^2 \kappa^2}+\phi_0^2 \kappa^2 t^5 t_0+c_2+\right.\nonumber\\
&+&\frac{5 t^4 t_0^2}{6 \phi_0^2 \kappa^2}-\frac{5}{24} \phi_0^2 \kappa^2 t^4 t_0^4-
-\frac{\lambda_0 t^4}{t_0^4}-\frac{\lambda_0 t^3}{2 t_0^3}-\frac{\lambda_0 t^2}{3 t_0^2}-\nonumber\\
&-&\frac{\lambda_0 t}{4 t_0}-\frac{1}{2} \lambda_0 \ln [t]+\frac{\lambda_0 t^5 \ln[t]}{t_0^5}-\frac{3}{2} \lambda_0\ln[t_0-t]-\nonumber\\
&-&\left.\frac{\lambda_0 t^5 \ln[t_0-t]}{t_0^5}\right),
\eea
\bea
V&=&
-\frac{\phi_0^2 \left(4 t^3+40 t^2 t_0-19 t t_0^2+3 t_0^3\right)}{2  t^2 (t-t_0)^3}-\nonumber\\
&-&\frac{4 t_0 \left(-3 c_1 \kappa^2 t^3+2 t_0 (3 t+t_0)\right)}{2 \kappa^2 t^2 (t-t_0)^3}+\nonumber\\
&+&
\frac{\phi_0^2 \lambda_0  \left(-60 t^5+30 t^4 t_0+10 t^3 t_0^2-15 t^2 t_0^3-t t_0^4+t_0^5\right)}{5 t^4 (t-t_0)^4 t_0^3}+\nonumber\\
&+&
\frac{60\phi_0^2 \lambda_0 t  (\ln[t]-\ln[t_0-t])}{5  (t-t_0)^3 t_0^4}.
\eea

In this case, we have  written the expression as a function of time, as we have the uncertainty noted above. 

In conclusion, it can be seen that the method works not only for monotonic  functions, but also for some non-monotonic Hubble  functions that have to be carefully considered. However problems  may arise by converting the functions $V$ and $\varepsilon$ in terms of the scalar field. In this process,  uncertainty on the behavior can arise. The interest  to derive  suitable Hubble rate forms (not necessarily monotonic) is related to the necessity to obtain   realistic dark energy models. In fact,   $H(t)$ decreasing with time is generally related to  the evolution  of  matter dominated  epochs and  large scale structure formation. On the other hand, $H(t)$  increasing with time can be related to inflationary and dark energy behaviors. Clearly,   special care has to be devoted to singularities at finite which could come out, as above, in this reconstruction process.


\section{Dynamical system analysis and critical points}

Let us rewrite the cosmological Friedmann  equations  in terms of an  autonomous dynamical system. The approach
   consists of two parts: the Friedmann Eqs. (\ref{LagS3}) and (\ref{LagS4}) and equations specifying
the model, for example  (\ref{eqq1}) and (\ref{eqq2}). It is convenient to work with the new
variables
\be \label{nv1}
x=\frac{\dot{\phi}}{H},\;\;\;y=\dot{\varepsilon}H,\;\;z=\frac{V}{H^2},\;\;\mu=\frac{\dot{H}}{H^2}.\ee
The prime means the derivative with respect to $\ln{a}$.

After some algebra, we obtain the following dynamical system
\bea \label{avt1}
0&=&-\frac{3}{\kappa^2}+\frac{1}{2}x^2+\lambda x^2+z+24 y,\nonumber\\
0&=&\frac{1}{\kappa^2}\left(2\mu+3\right)+\frac{1}{2}x^2-z-8y'-8y\mu-16y,\nonumber\\
y'&=&y(b_2 x +\frac{x'}{x}+2\mu),\\
z'&=&b_1z\,x-2z\mu-\frac{b_1\Lambda x}{H^2},\nonumber\\
\lambda'&=&b_3(\lambda-\lambda_1)x\,,\nonumber
\eea
where the first equation is a constraint for the other 4 equations.
For the sake of simplicity, we have considerd the case  $\Lambda=0$.
As usual, the critical points at finite are given by the condition

\be x'=y'=z'=\lambda'=0\,.\ee 

The various situations are

\begin{itemize}
\item{\bf A}: $(x,y,z,\lambda)=(0,0,\frac{3}{\kappa^2},\lambda)$ or
$=(0,0,\frac{3}{\kappa^2},\lambda_1)$ (two points - any $\lambda$ or 
$\lambda=\lambda_1$
is constant).
For this point $\mu = 0$ and hence EoS parameter  is equal to minus one
($w=-1$). Thus, it  is a de Sitter space. In this point, the potential of the
scalar field dominates.

\item{\bf B}:
$(x,y,z,\lambda)$=$(-\frac{b_1}{\kappa^2(1+\lambda_1)},0,\frac{6\kappa^2(1+\lambda_1)-b_1(1+2\lambda_1)}{2\kappa^4(1+\lambda_1)^2},\lambda_1)$
($\lambda$ is constant).
For this point $\mu =-\frac{b_1^2}{2\kappa^2(1+\lambda_1)}$ and hence
EoS parameter  is equal: $w=\frac{b_1^2}{3\kappa^2(1+\lambda_1)}-1$.  Also in this
point, the scalar field potential dominates.

\item{\bf C}:
$(x,y,z,\lambda)=(\pm\frac{\sqrt{6}}{\sqrt{\kappa^2(1+2\lambda_1)}},0,0,\lambda_1)$
($\lambda$ is constant).
For this point $\mu =-\frac{3(1+\lambda_1)}{1+2\lambda_1}$ and hence EoS
parameter   is  $w=\frac{2(1+\lambda_1)}{1+2\lambda_1}-1$.  In this point the
kinetic energy of the scalar field dominates.

\item{\bf D}: For this point one gets complicated expressions for $x$ and $y$ ($x\ne
0,\;\;y\ne 0$), but $z=0$ and $\lambda=\lambda_1=const$
Futhermore, one gets:  $\mu =-\frac{x}{3b_2}$.

\end{itemize}

If $\Lambda$ is not zero (we have to add one
more equation $H'=\mu H$ to the system (\ref{avt1})), then the point {\bf A} does not change. However, it
appears a 
   new point  {\bf A1}  for which $H = \pm \frac{\kappa \sqrt{\Lambda}}{\sqrt{3}}$.
This means that at least one more point with all of the variables  not equal to zero
exists. Following this analysis of critical points, it is straightforward to study the
stability and attractor nature of cosmological solutions under discussion at finite and asymptotically. 

Finally we want to stress that  the presence of the Lagrange multiplier adds new critical points.
Moreover, we can get a different value $w$ for the same potential by the
modification of the constraint.


\section{Conclusions}

We  have examined a  string-inspired effective theory of gravity containing Gauss-Bonnet
invariant interacting with a scalar field where  a  Lagrange multiplier is inserted into the action.
The  Lagrange multiplier term can be related to the process of  string
compactification that leads to the effective 4D-action.
 Adopting a FRW metric,  the corresponding cosmological Friedmann equations
(\ref{LagS3}) and (\ref{LagS4}) can be
used to   define  the effective potential and the Lagrange multiplier (and the  function
$\varepsilon$ which is the coupling to the Gauss-Bonnet term).
Choosing a  form for the metric and the scalar field, one can easily approach  the
cosmological reconstruction in a sort of inverse scattering approach. In general, the presence of the Lagrange multiplier
 helps in the generation of new cosmological solutions or in changing  
the features of some known solutions. This fact could be extremely relevant from an observational point of view because the Lagrange multipliers could be related to natural "priors" in order to discriminate among concurrent models. Specifically, relating cosmographic observed parameters as $\{H_0, q_0, j_0, s_0, \Omega_M, \Omega_{\Lambda}\}$ to some  Lagrange multiplier could result a useful tool to restrict the range of models physically viable   (see Ref.\cite{review} and references therein for a detailed discussion on this point).

In the models that we have considered, the Lagrange multiplier can be interpreted as the presence of
some dust fluid (matter without pressure) that affects the evolution of the cosmological system. In this case, the Lagrange multiplier can
help to realize the  transitions between  matter dominance and
dark energy era (and vice-versa) without imposing unnatural fine-tunings.
The detailed study of the theory  (\ref{LagS1}) with the exponential choice for
the potential, the
Lagrangian multiplier and the function $\varepsilon$ has been  presented. Two forms of
scalar field: $\phi \sim t$ and $\phi \sim
\ln{t}$ are discussed. The comparison with the case of no Lagrange multiplier
is considered. The presence of the Lagrange multiplier gives rise to  new cosmological
solutions due to the change in the dynamical system. For example, in the case  of the  scalar field  with  the 
logarithm form in  time, the theory provides only one cosmological solution. In
 presence of the Lagrange multiplier, the number of cosmological solutions with different features
increase. In particular, (ghost-free) phantom
cosmology with canonical scalar  easily emerges as solution. However, the situation can be more complicated since the Hubble rate $H$ is not necessarily a monotonic function of time. In some phantom models, dark energy grows with time and so does $H$ at late times. On the other hand, $H$ can decrease in matter-dominated eras so non-monotonic behaviors have to be considered.
Above we gave also an example in this sense.

The extension to other modified gravities (teleparallel, Horava-Lifshitz, $F(G)$,
non-local, etc \cite{review1}) via the introduction of the Lagrange multiplier 
can be easily accomplished in this framework. 
These topics will be discussed elsewhere.

\appendix
\section{ The Lagrange Multiplier Method and Noether Symmetries}

A more detailed discussion on the Lagrange Multiplier Method is necessary. A Lagrange multiplier is not introduced {\it ad hoc} in the dynamics but it is related to constraints of the theory related to  symmetries and conservation laws. Here we will sketch the Noether Symmetry Approach 
(see Ref.\cite{N} for a recent review) showing  that the reduction of dynamics induced by the Lagrange multipliers is always related to the search for symmetries \cite{basilakos,defelice}. 
In particular, the  form of the Lagrange multiplier derives from conservation laws and then it has always a physical meaning.

We have used a FRW metric, so that the Einstein
Eqs. (\ref{LagS3}) and (\ref{LagS4}) reduce to ordinary differential equations. The same equations can  be derived from the action (\ref{LagS1}) where 
 the  Lagrangian becomes  point-like after introducing in it a FRW metric.  The configuration space is
a  {\it minisuperspace}
where
the Lagrangian coordinates are  the scale factor $a$ and 
the scalar field $\p$,
with  the velocities $\dot{a},\dot{\p}$  \cite{N}. 
In this case, the dimension of the space is  2 but there are cases in which the minisuperspace dimension 
can  be larger (for instance a Bianchi universe with anisotropies \cite{marmo}
 or  dynamical systems with more than one scalar field \cite{lambiase}). 

With these considerations in mind, let ${\l}(q^{i}, \dot{q}^i)$ be a Lagrangian, independent of time
and nondegenerate, i.e. 
\beq
\label{01}
\frac{\pa {\l}}{\pa t}=0\,;\;\;\;\;\;\;\;
\mbox{det}H_{ij}\eqdef
\mbox{det}
\left|\left| \frac{\pa^2 {\l}}{\pa \dot{q}^i\pa\dot{q}^j}\right|\right|\neq 0\,,
\eeq
where det $H_{ij}$ is the Hessian determinant. In standard problems of analytical mechanics ${\l}$  has the form
\beq
\label{02}
{\l}=T({\bq},\dot{\bq})-V({\bq})\;,
\eeq
where $T$ and $V$ are the kinetic  and potential energy.
$T$ is a positive definite quadratic form in $\dot{\bq}$.
Associated with ${\l}$ is the energy function  
\beq
\label{03}
E_{\l}\equiv\frac{\pa {\l}}{\pa \qd^{i}}\qd^i-{\l}\,,
\eeq
which is  the total energy $T+V$. It is worth noticing that ${\l}$ can be  more complicated 
than  (\ref{02})
and $E_{\l}$ is  a constant of the motion
 called "energy" also in general cases.
In the Lagrangian formalism, we have to consider only transformations which
are
point-transformations.  Any invertible and smooth transformation of the
positions $Q^{i}=Q^{i}({\bq})$ induces a transformation of the 
velocities of the form
\beq
\label{04}
\dot{Q}^i({\bq})=\frac{\pa Q^i}{\pa q^j}\qd^j\;.
\eeq
The matrix ${\cal J}=|| \pa Q^i/\pa q^j ||$ is the Jacobian of the
transformation on the positions and it is assumed to be nonzero.
The Jacobian ${\cal J}'$ of the induced transformation is easily derived
and ${\cal J}\neq 0\rightarrow {\cal J}'\neq 0$. 
A point transformation $Q^{i}=Q^{i}(\bq)$ can depend on one
(or more than one) parameter. 
Let us assume that a point transformation depends on a parameter $\epsilon$,
i.e. $Q^{i}=Q^{i}(\bq,\epsilon)$, and that it gives rise to a 
one--parameter Lie group. For infinitesimal values of $\epsilon$, the 
transformation is then generated by a vector field:
for instance, as well known, 
$\pa/\pa x$ represents a translation along $x$ axis, 
$x(\pa/\pa y)-y(\pa/\pa x)$ is a rotation around $z$ axis and so on.
In general, an infinitesimal point transformation is represented by a 
generic vector field on $Q$
\beq
\label{04b}
{\bf X}=\alp^i({\bq})\frac{\pa}{\pa q^i}\;.
\eeq
The induced transformation (\ref{04}), considering also velocities,  is then represented by
\beq
\label{05}
{\bf X}^c=\alp^{i}({\bq})\frac{\pa}{\pa q^{i}}+
\left(\frac{d}{dt}\alp^{i}({\bq})\right)\frac{\pa}{\pa \qd^j}\;.
\eeq
${\bf X}^c$ is called the "complete lift" of ${\bf X}$ 
\cite{arnold}. From now on, we will drop the suffix $^c$ but clearly we refer to a complete lift.
A function $f(\bq, \bqd)$ is invariant under a transformation represented
by ${\bf X}$ if
\beq
\label{06}
L_{{\bf X}}f\eqdef\alp^{i}({\bq})\frac{\pa f}{\pa q^{i}}+
\left(\frac{d}{dt}\alp^{i}({\bq})\right)\frac{\pa f}{\pa \qd^j}=0\;,
\eeq
where $L_{{{\bf X}}}f$ is the Lie derivative of $f$. In particular,
if $L_{{{\bf X}}}{\l}=0$, ${\bf X}$ is said to be a 
{\it symmetry} for the
dynamics derived by ${\l}$. 

Let us consider  now a Lagrangian ${\l}$ and its Euler-Lagrange equations
\beq
\label{07}
\frac{d}{dt}\frac{\pa {\l}}{\pa\qd^{j}}-\frac{\pa {\l}}{\pa q^{j}}=0\,.
\eeq
Let us consider also a vector field of the form (\ref{05}). Contracting 
(\ref{07}) with the $\alpha^{i}$'s gives
\beq
\label{06a}
\alp^{j}\left( \frac{d}{dt}\frac{\pa {\l}}{\pa \qd^j}-
\frac{\pa {\l}}{\pa q^j}\right)=0\,.
\eeq
Being
\beq
\label{06b}
\alp^{j}\frac{d}{dt}\frac{\pa {\l}}{\pa \qd^j}=
\frac{d}{dt}\left(\alp^j\frac{\pa {\l}}{\pa \qd ^j}\right)-
\left(\frac{d \alp^j}{dt}\right)\frac{\pa {\l}}{\pa \qd ^j}\,,
\eeq
from (\ref{06a}), we obtain
\beq
\label{08}
\frac{d}{dt}\left(\alp^{i}\frac{\pa {\l}}{\pa \qd^i} \right)=L_{{\bf X}}{\l}\,.
\eeq
The immediate consequence is the  
{\it Noether Theorem}: {\it If $L_{{\bf X}}{\l}=0$, then the function 
\beq
\label{09}
\Sigma_{0}=\alp^{i}\frac{\pa {\l}}{\pa \qd^i} \,,
\eeq
is a constant of motion.}

\vspace{3. mm}

\noindent Eq.(\ref{09}) can be expressed independently of
 coordinates as a contraction of ${\bf X}$ with Cartan one--form 
\beq
\label{09a}
\theta_{\l} \eqdef \frac{\pa {\l}}{\pa \qd^i}dq^i \; .
\eeq
For a generic vector field $ {\bf Y} = y^i \pa / \pa x^i $, and one--form  
$\beta
= \beta_i d x^i $, we have by definition $ i_{\bf Y} \beta = y^i \beta_i $. 
Thus 
Eq.(\ref{09}) can be written as
\beq
\label{09b}
i_{{\bf X}} \theta _{\l} = \Sigma_0  \; .
\eeq
Under a  point--transformation,   the vector field ${\bf X}$ becomes
\beq
\label{09c}
{\bf X}' = (i_{\bf X} d Q^k) \frac{\pa}{\pa Q^k} + 
     \left( \frac{d}{dt} (i_x d Q^k)\right) \frac{\pa}{\pa \dot{Q}^k} \; .
\eeq 
We see that ${\bf X}'$ is still the lift of a vector 
field defined on the "positions"
space only. If ${\bf X}$ is a symmetry and we 
choose a point transformation in such a way
that 
\beq
\label{010}
i_{\bf X} dQ^1 = 1 \; ; \;\;\; i_{\bf X} dQ^i = 0  \;\;\; i \neq 1 \; ,
\eeq
we get 
\beq
\label{010a}
{\bf X}' = \frac{\pa}{\pa Q^1} \;;\;\;\;\;  \frac{\pa {\l}}{\pa Q^1} = 0 \; .
\eeq
Thus $Q^1$ is a cyclic coordinate and the dynamics can be reduced according 
to  well known procedures \cite{arnold,cimento}. The change of coordinates defined by (\ref{010}) is not unique. Usually
a clever choice is very important. However, more than one symmetry can exist and the degree of dynamical reduction depends on the number of symmetries \cite{N}.

Let us now assume that ${\l}$ is a canonical Lagrangian (e.g. of the form (\ref{02})).
As ${\bf X}$ is of the form (\ref{09c}), $L_{\bf X}{\l}$ will be a 
homogeneous polynomial of second degree in the velocities plus a 
inhomogeneous term in the $q^{i}$. Since such a polynomial
 has to be identically zero,
each coefficient must be independently zero. If $n$ is the dimension of the 
configuration space, we get $\{1+n(n+1)/2\}$ partial differential
equations (PDE). The system is overdetermined, 
therefore, if a solution exists,
it will be expressed in terms of integration constants instead of boundary 
conditions. Usually, ${\l}$ is fixed and one can ask for
solutions of a specific PDE system.
In particular, a  Lagrangian  containing 
 some undefined functions (e.g.  couplings $\omega(\phi)$, $\varepsilon(\phi)$ and  potentials  $V(\phi)$)
is  a {\it class of Lagrangians} where the single element is determined by the solution of the
 PDE systems. 
In other words,  the Noether symmetry   selects the 
 functions which assign the model. 

The above discussion shows that, given a dynamical system, it is
always possible to search for Noether symmetries. If they exist,
 the dynamics can be specified and reduced by a 
change of variables since one (or more than one) coordinate becomes
cyclic. If we are able to integrate such a new dynamics, the  problem
is to invert the solution in order to get the evolution in the
previous variables \cite{basilakos,defelice}. 
Conversely, the existence of  Noether symmetries is able to select the
form of scalar field potentials  and  couplings. In other words,
given a class of models, the symmetries are able to select some of them,
which, in principle,   are of physical interest.  

The Lagrange multiplier method is related to this approach. In other words, imposing the existence of  
Lagrange multipliers means to select models where symmetries exist because the Lagrangian functions become canonical. The example below shows the strict relation between the Lagrange multipliers and the Noether symmetries.

Let us consider the  simple case of $f(R)$ gravity defined by the action
\beq
\label{3.7.1}
 S=\vol f(R)\,,
\eeq
where, as usual, $R$ is the Ricci scalar.  This case is very interesting since not only the coupling and the potential are unspecified but the whole Lagrangian is not given a priori. Imposing, as above, the  FRW metric, we  can write
\beq
\label{3.7.3}
S=\int {\l}(a,\dot{a},R,\dot{R})dt\,,
\eeq
considering $a$ and $R$ as canonical variables. 
Such a  position is
arbitrary  since $R$ depends on $a$,  $\dot{a}$ and $\ddot{a}$ so (\ref{3.7.3}) is not a true canonical Lagrangian.
To remove this ambiguity, one can use a  Lagrange multiplier $\lambda$ and search for Noether symmetry related to it. We have
\bea
\label{3.7.4}
S&=&2\pi^{2}\int dt\times\\
&\times&\left\{f(R)a^{3}-\lambda\left[
R+6\left(\frac{\ddot{a}}{a}+\frac{\dot{a}^{2}}{a^{2}}+\frac{k}{a^{2}}\right)
\right]\right\}\,.\nonumber
\eea
To determine $\lambda$, we have to vary the action with respect 
to $R$, that is
\beq
\label{3.7.5}
a^{3}\frac{df(R)}{dR}\delta R -\lambda\delta R=0\,,
\eeq
from which we get, like in Sec.II,  the functional form of the Lagrange multiplier, that is 
\beq
\label{3.7.6}
\lambda =a^{3}f'(R)\,.
\eeq
Substituting into (\ref{3.7.4}) and integrating by parts, we obtain the point-like
Lagrangian
\bea
\label{3.7.7}
{\l}&=&a^{3}\left[f(R)-Rf'(R)\right]+6\dot{a}^{2}af'(R)+\nn
&+&6a^{2}\dot{a}\dot{R}f''(R)-akf'(R)\,.
\eea
Then the cosmological equations of motion are
\bea\label{pippo}
&&\left(\frac{\ddot{a}}{a}\right)f(R)'+
2\left(\frac{\dot{a}}{a}\right)f(R)''\dot{R}+\\ &+&f(R)''\ddot{R}+f(R)'''\dot{R}^{2}-
\frac{1}{2}[Rf(R)'+f(R)]=0\,,\nonumber
\eea
and
\beq
\label{pluto}
R=-6\left(\frac{\ddot{a}}{a}+\frac{\dot{a}^{2}}{a^{2}}+\frac{k}{a^{2}}\right)\,,
\eeq
where, as above,  the Lagrange multiplier gives one of the equations of motion.
The energy constraint is
\bea\label{paperino}
&&6\dot{a}^{2}af'(R)-a^{3}\left[f(R)-Rf'(R)\right]+\\&+&
6a^{2}\dot{a}\dot{R}f''(R)+akf'(R)=0\,.\nonumber
\eea
The symmetry generator is defined on the tangent bundle 
$TQ(a,\dot{a},R,\dot{R})$ and it is
\beq
\label{3.7.8}
{\bf X}=\alpha(a,R)\frac{\pa}{\pa a}+\beta(a,R)\frac{\pa}{\pa R}+
\frac{d\alpha}{dt}\frac{\pa}{\pa\dot{a}}
+\frac{d\beta}{dt}\frac{\pa}{\pa\dot{R}}\,,
\eeq
while the Noether condition $L_{\bf X}{\l}=0$ produces the system
\beq
\label{3.7.9}
 f'(R)\left[\alpha+2a\frac{\pa \alpha}{\pa a}\right]+
af''(R)\left[\beta+a\frac{\pa \beta}{\pa a}\right]=0 \,,
\eeq
\beq
\label{3.7.10}
 a^{2}f''(R)\frac{\pa \alpha}{\pa R}=0\,,
\eeq
\bea
\label{3.7.11}
2f'(R)\frac{\pa \alpha}{\pa R}+f''(R)\left[
2\alpha+a\frac{\pa \alpha}{\pa a}+a\frac{\pa \beta}{\pa R}\right]
&+&a\beta f'''(R)=\nn
&&=0\,,
\eea
\beq
\label{3.7.12}
 3\alpha\left[f(R)-Rf'(R)\right]-a\beta Rf''(R)=0\,,
\eeq
\beq
\label{3.7.13}
\alpha f'(R)+a\beta f''(R)=0\,.
\eeq
From (\ref{3.7.10}), 
we have that $\alpha$ is a function of $a$ only, if we want
to avoid trivial cases (i.e. we want  $f''(R)\neq 0$).
The symmetry is given by the functions
\beq
\label{3.7.14}
\alpha=\frac{\beta_{0}}{a}\,,\;\;\;\beta=-2\beta_{0}\frac{R}{a^{2}}\,,\;\;\;
f(R)=f_{0}R^{3/2}\,,
\eeq
which solve the above system;  $\beta_{0}$ and $f_{0}$ are integration constants.
The new induced variables can be
\beq
\label{3.7.15}
w=a^{2}R\,,\;\;\;\;\;z=\frac{a^{2}}{2\beta_{0}}\,,
\eeq
from which the Lagrangian (\ref{3.7.7}) becomes
\beq
\label{3.7.16}
\tilde{\l}=\frac{9}{2}\beta_{0}\frac{\dot{z}\dot{w}}{\sqrt{w}}-9k\sqrt{w}-
\frac{1}{2}\sqrt{w^{3}}\,,
\eeq
which can be rewritten in the form
\beq
\label{3.7.17}
\tilde{\l}=9\beta_{0}\dot{z}\dot{y}-9ky-\frac{1}{2}y^{3}\,,
\eeq
using $y=\sqrt{w}$. Dynamics is then described from the  equations
\beq
\label{3.7.18}
\ddot{y}=0\,,\;\;\;\;\mbox{from which }\;\;\dot{y}=\dot{y}_{0}=\Sigma_{0}\,,
\eeq
\beq
\label{3.7.19}
9\beta_{0}\ddot{z}+9k+\frac{3}{2}y^{2}=0\,,
\eeq
\beq
\label{3.7.20}
 9\beta_{0}\dot{y}\dot{z}+9ky+\frac{1}{2}y^{3}=0\,,
\eeq
whose solutions are
\beq
\label{3.7.21}
y(t)=\dot{y}_{0}t+y_{0}\,,
\eeq
\beq
\label{3.7.22}
z(t)=c_{4}t^{4}+c_{3}t^{3}+c_{2}t^{2}+c_{1}t+c_{0}\,,
\eeq
with
\bea
\label{3.7.23}
c_{4}&=&-\frac{\dot{y}_{0}^{2}}{72\beta_{0}}\,,\;\;
c_{3}=-\frac{\dot{y}_{0}y_{0}}{3\beta_{0}}\,,\nn
c_{2}&=&-\frac{y_{0}^{2}}{12\beta_{0}}-\frac{k}{2}\,,\;\;
c_{1}=\dot{z}_{0} \,,\;\;
c_{0}=z_{0}\,.
\eea
The energy condition (\ref{3.7.20}) gives the relation among the initial data \cite{defelice}.
Going back to the physical variables, we have
\beq
\label{3.7.24}
a(t)=\pm\sqrt{d_{4}t^{4}+d_{3}t^{3}+d_{2}t^{2}+d_{1}t+d_{0}}\,,
\eeq
where the $d_{i}$'s are  the $c_{i}$'s multiplied by $2\beta_{0}$.
The $R$ variable is actually the Lagrange multiplier which gives
\beq
\label{3.7.25}
R=\frac{(\dot{y}_{0}t+y_{0})^{2}}{d_{4}t^{4}+d_{3}t^{3}+d_{2}t^{2}+
d_{1}t+d_{0}}\,,
\eeq
and then the  cosmological equations (\ref{pippo}), (\ref{pluto}) and (\ref{paperino}) are fully satisfied. In conclusion, by imposing the Lagrange multiplier we get a canonical dynamics and the consequent  existence of the Noether symmetry allows its integration. In summary, the dynamical system results both canonical and integrable. More complicated cases are discussed  in \cite{N,Odin21}.

\section*{Acknowledgments.}

This work has been supported by  project 2.1839.2011 of Min. of Education and
Science (Russia) and LRSS project  224.2012.2 (Russia). S.C. is supported by INFN (Iniziative specifiche NA12 and OG51).
The work by SDO is supported in part by MICINN (Spain),  project
FIS2010-15640 and by AGAUR (Generalitat de Ca\-ta\-lu\-nya),
contract 2009SGR-994.

\end{document}